\documentclass{iaus}
\usepackage{graphicx}

\title[Theories of the initial mass function] 
{Theories of the initial mass function}

\author[Patrick Hennebelle \& Gilles Chabrier]   
{Patrick Hennebelle$^1$
 \and Gilles Chabrier$^2$}

\affiliation{$^1$Laboratoire de radioastronomie, Ecole normale sup\'erieure and Observatoire de Paris,  \\ UMR CNRS 8112 24 rue Lhomond, 75231 Paris Cedex 05, France  \\ email: {\tt patrick.hennebelle@ens.fr} \\[\affilskip]
$^2$CRAL, Ecole normale sup\'erieure de Lyon, UMR CNRS 5574 \\ Univesrit\'e de Lyon, 69364 Lyon Cedex 07, France \\email: {\tt chabrier@ens-lyon.fr}}

\pubyear{2008}
\volume{xxx}  
\pagerange{119--126}
\jname{Title of your IAU Symposium}
\editors{A.C. Editor, B.D. Editor \& C.E. Editor, eds.}
\begin{document}

\maketitle

\begin{abstract}
We review the various theories which have been proposed along the years
to explain the origin of the stellar initial mass function. We pay particular 
attention to  four models, namely  the competitive accretion and the 
theories based respectively  on stopped accretion, MHD shocks and 
turbulent dispersion. In each case, we derive the main 
assumptions and calculations that support each theory and stress their respective
successes and failures or difficulties.  
\keywords{STARS: formation, ISM: clouds, gravitation, turbulence}
\end{abstract}

\firstsection 
\section{Introduction}
Stars are the building blocks of our universe and understanding 
their formation and evolution is one of the most important problems 
of astrophysics. Of particular importance is the problem 
of the initial mass function of stars (Salpeter 1955, Scalo 1986,
Kroupa 2002, Chabrier 2003) largely 
because the star properties, evolution and influence on the surrounding interstellar medium
  strongly depend on their masses. 
 It is generally found that 
the number of stars per logarithmic bin of masses,
$d N / d \log M$,  can be described
by a lognormal 
distribution below 1 $M_\odot$,  peaking at about $\simeq$0.3 $M_\odot$, and a power-law of slope 
$-1.3$ for masses between 1 and 10 $M_\odot$\footnote{ It should be kept in mind that the {\it initial} mass function is measured only up to
about 8 $M_\odot$, in young stellar clusters, and is inferred only indirectly for larger masses (see e.g. Kroupa 2002)} (e.g. Chabrier 2003).
It should be stressed that the IMF of more massive stars is extremely 
poorly known.

Several theories have been proposed to explain the origin of the IMF, invoking 
various physical processes that we tentatively classify in 
four categories: theories based on recursive fragmentation or pure 
gravity (e.g. Larson 1973), theories based on pure statistical argument, invoking 
the central limit theorem (e.g. Zinnecker 1984, Elmegreen 1997,  Adams \& Fatuzzo 1996), theories based on accretion
  and, finally, theories invoking the initial Jeans mass in a fluctuating environment.  We will focus on the two latter
  ones, which appear to be favored in the modern context of star formation.


\section{Theories based on accretion}

\subsection{Competitive accretion}

The theory of competitive accretion has been originally proposed
by Zinnecker (1982) and Bonnell et al. (2001).
It has then been used to interpret the series of numerical simulations 
similar to the ones performed by Bate et al. (2003).

The underlying main idea is that the accretion onto the star is directly linked
to its mass in such a way that massive stars tend to accrete more efficiently
and thus become  disproportionally more massive than the low mass stars.
The accretion rate is written as:
\begin{eqnarray} 
\dot{M}_* = \pi \rho V_{rel} R_{acc}^2,
\label{accret}
\end{eqnarray}  
where $\rho$ is the gas density, $V_{rel}$ is the relative velocity between the star and the gas
 while  $R_{acc}$ is the accretion radius. 
Bonnell et al. (2001) consider two situations, namely the cases where the gravitational
potential is dominated by the gas or by the stars.

\subsubsection{Gas dominated potential} 
Let  $R$ be the spherical radius, $\rho$ the gas density and $n_*$ the number 
density of stars. 
In the gas dominated potential case, Bonnell et al. (2001) assume, following 
Shu (1977), 
that the gas density profile is proportional to $R ^{-2}$. They further 
assume that $n_* \propto R^{-2}$. The accretion radius, $R_{acc}$ is 
assumed to be equal to the tidal radius given 
by 
\begin{eqnarray} 
R_{tidal} \simeq 0.5 \left( {M_*\over M_{enc}} \right)^{1/3}  R,
\label{Rtidal}
\end{eqnarray}  
where $M_{enc}$ is the mass enclosed within radius $R$.
This choice is motivated by the fact that a fluid particle located at a distance
from a star smaller than $R_{tidal}$  is more sensitive to the star than to the cluster potential and will thus be accreted onto this 
star.

The mass of gas within a radius $R$, $M(R)$, is proportional to $R$ (since $\rho \propto
 R^{-2}$). The infall speed is about 
 $V_{in} \simeq \sqrt{G M(R) /R}$, and, assuming that the stars are virialized,
 one gets $V_{rel} \simeq V_{in}$.
The number of stars, $dN_*$, located between  $R$ and $R+dR$ is given by the relation 
$dN_* = n_*(R) \times 4 \pi R^2 dR \propto dR$. Thus 
eqn~\ref{accret} combined with the expression of 
$V_{rel}$ and $R_{acc}$ leads to the relation $\dot{M}_* \propto (M_*/R)^{2/3}$ and, 
after integration, 
 $M_* \propto R^{-2}$, $R \propto M_*^{-1/2}$. Consequently, we obtain:
\begin{eqnarray} 
dN \propto M_*^{-3/2} dM_*.
\label{dNacc1}
\end{eqnarray}  
Even though the mass spectrum is too shallow compared to the fiducial IMF, $dN/dM \propto M^{-2.3}$, it is interesting to see that this power law 
behaviour can be obtained from such a simple model. Note, however, that 
the model implies that stars of a given mass are all located in the same 
radius, which points to a difficulty of the model.

\subsubsection{Star dominated potential} 
When the potential is dominated by the stars located in the centre of the cloud,  
the density is given by $\rho \propto R^{-3/2}$, which corresponds to the expected density 
distribution after the rarefaction wave has propagated away
(Shu 1977). The velocity is still assumed to be $V_{rel} \propto R^{-1/2}$. 
The accretion radius is now supposed to correspond to the Bondi-Hoyle radius 
as the gas and the star velocities are no longer correlated.
This leads to
$\dot{M}_* = \pi \rho V_{rel} R_{BH}^2$ where $R_{BH} \propto M_* / V_{rel}^2$. 
It follows
\begin{eqnarray} 
\dot{M}_* \propto M_*^{-2}.
\label{dMacc2}
\end{eqnarray} 
One can then show (Zinnecker 1982, Bonnell et al. 2001) that under reasonable assumptions,
 $dN \propto M_*^2 dM_*$. This estimate is in better agreement with the 
Salpeter exponent, although still slightly too shallow. These trends 
seem to be confirmed by the simulations performed by Bonnell et al. (2001)
which consists in distributing 100 sink particles in a cloud of total mass
about 10 times the total mass of the sinks initially (their Fig. 3). 

\subsubsection{Difficulties of the competitive accretion scenario}
As obvious from the previous analytical derivations, finding an explanation for 
the Salpeter exponent with the competitive accretion scenario appears to be difficult, 
even though numerical simulations (as the ones presented in Bonnell et al. 2001) seem to successfully achieve this task.
However, although the IMF exponent 
is close to the Salpeter one in the star dominated potential case,  this scenario entails by construction the Bondi-Hoyle accretion, which  is at least a factor ~3 lower than the mass infall rate resulting from gravitational collapse at the class O and I stages and leads to too long accretion times compared with observations (Andr\'e et al. 2007, 2009). Another difficulty
of this scenario is that it does not explain the peak of the IMF which 
might be related to the Jeans mass (see \S~\ref{jeans}). Finally, 
it is not clear that competitive accretion can work 
in the case of non-clustered star formation, for which the gas density
is much too small. As no evidence for substantial IMF variation  among different regions
has yet been  reported, this constitutes a  difficulty for this model as a general model for star formation.
Perhaps this scenario applies well to the formation of massive stars.

\subsection{Stopped accretion}
The principle of this type of models is to assume that the accretion of gas onto 
the stars or the dense cores is a non-steady process, stopped because of either the finite reservoir of mass
or the influence of an outflow which sweeps up the 
remaining gas within the vicinity of the accreting protostar.

The first studies were performed by Silk (1995) and Adams \& Fatuzzo (1996). 
They first relate the mass of the stars to the physical parameters of the cloud such as
sound speed and rotation and then assume that an outflow whose properties 
are related to the accretion luminosity stops the cloud collapse. 
Using the Larson (1981) relations, they can link all these parameters to 
the clump masses. Since the mass spectrum of these latter is known (e.g. Heithausen et al. 1998),
they infer the IMF.

A  statistical approach has been carried out by Basu \& Jones (2004).
These authors assume that the dense core distribution is initially 
lognormal, justifying it by the large number of processes that control their 
formation (and invoking the central limit theorem). Then, they argue that 
the cores grow by accretion and postulate that the accretion rate is 
simply proportional to their mass, 
$\dot{M} = \gamma M \rightarrow M(t)=M_0 \exp(\gamma t)$,
leading  to $\log M = \mu = \mu_0 + \gamma t$. Finally, they assume that accretion 
is lasting over a finite period of time given by 
$f(t)=\delta \exp(-\delta t)$.
The star mass distribution is thus obtained by summing over the 
accretion time distribution.
\begin{eqnarray}
f(M) &=& \int _0 ^\infty { \delta \exp(-\delta t) \over \sqrt{2 \pi} \sigma_0 M } \exp \left( -{ (\ln M - \mu_0 -\gamma t)^2\over 2 \sigma_0^2}  \right) dt \\
&=& {\alpha \over 2} \exp( \alpha \mu_0 + \alpha^2 \sigma_0^2/2) M^{-1-\alpha} {\rm erf} \left( {1 \over \sqrt{2} } (\alpha \sigma_0 - { \ln M - \mu_0 \over \sigma_0})\right),
\nonumber
\end{eqnarray}
where $\alpha=\delta/\gamma$ and $\sigma_0$ characterizes the width of the initial dense
core distribution. As  $\delta$ and $\alpha$ are controlled by the same types 
of processes, their ratio is expected to be of the order of  unity
and thus $f(M)$ exhibits a powerlaw  behaviour close to the fiducial IMF. 

In a recent study, Myers (2009) develops similar ideas in  more details, 
taking  into account the accretion coming from the surrounding 
background. Adjusting two parameters, he reproduces quite nicely the observed 
IMF (his figure 5).

A related model has also been developed by Bate \& Bonnell (2005) based on an 
idea proposed by Price \& Podsiadlowski (1995). They consider objects
that form by fragmentation within a small cluster and are ejected 
by gravitational interaction with the other fragments, which stops the accretion process. 
Assuming a lognormal accretion rate and an exponential probability of being ejected, 
these authors construct a mass distribution that can fit the IMF for some choices of parameters. 

In summary, the stopped accretion scenario presents interesting ideas and, providing (typically 2 or more) adequate adjustable parameters, 
can reproduce reasonably well the IMF. However, the very presence of such parameters, which characterizes our inability to precisely determine the processes that halt accretion,
illustrates the obvious difficulties of this 
class of models, and their lack of predictive power and accuracy. 

\section{Gravo-turbulent theories}
\label{jeans}

While in the accretion models, turbulence is not determinant, 
it is one of the essential physical processes for the two theories presented in this 
section, although the  role it plays differs in both models, as shown below. The theories proposed along this line
seemingly identify cores or {\it pre-cores} and are motivated 
by the strong similarity between the observed CMF and the IMF (e.g. Andr\'e et al. 
2010).

The first theory which combined turbulence and gravity was
proposed by Padoan et al. (1997). In this paper, the authors consider
a lognormal density distribution - density PDF computed from
 numerical simulations (e.g. V\'azquez-Semadeni 1994, Kritsuk et al. 2007, Schmidt et al. 2009, Federrath et al. 2010) 
are indeed nearly lognormal - and select the regions of the flow which 
are Jeans unstable.
By doing so, they get too stiff an IMF (typically 
$dN/ dM \propto M^{-3}$) but nevertheless find a lognormal behaviour at small 
masses, a direct consequence of the lognormal density distribution,
 and a powerlaw one at large masses.

\subsection{Formation of cores by MHD shocks}
The idea developed by Padoan \& Nordlund (2002) is slightly different. 
These authors consider a compressed layer formed by ram pressure in a weakly 
magnetized medium. They assume that the magnetic field is parallel to 
the layer and thus perpendicular to the incoming velocity field. 
The postshock density, $\rho_1$, the thickness of the layer, $\lambda$,
and the postshock magnetic field, $B_1$, can be related to the 
Alfv\'enic Mach numbers, ${\cal M}_a=v/v_a$ 
($v$ is the velocity and $v_a$ the Alfv\'en speed),
  and preshocked quantities, $\rho_0$ and $B_0$ according to the shock conditions:
\begin{eqnarray}
\rho_1/\rho_0 \simeq {\cal M}_a \; , \; \lambda / L \simeq {\cal M}_a^{-1} \; , \; B_1/B_0 \simeq {\cal M}_a \; ,
\label{eq_pn1}
\end{eqnarray}
where $L$ is the scale of the turbulent fluctuation. Note that for classical 
hydrodynamical isothermal shocks, the jump condition is typically $\propto {\cal M}^2$. 
The dependence on ${\cal M}_a$ instead of ${\cal M}^2$ stems from the 
magnetic pressure which is quadratic in $B$. As we will see, this 
is a central assumption of this model. 

The typical mass of this perturbation is expected to be
\begin{eqnarray}
m \simeq \rho_1 \lambda^3 \simeq \rho_0 {\cal M}_a \left( {L \over {\cal M}_a } \right)^3 \simeq \rho_0 L^3  {\cal M}_a^{-2}.
\label{eq_pn2}
\end{eqnarray}
As the flow is turbulent, the velocity distribution depends on the scale 
and $v \simeq L^\alpha$, with $\alpha=(n-3)/2$,
 $E(k)\propto k^{-n}$ being the velocity powerspectrum{\footnote{$n$ is denoted $\beta$ in Padoan \& Nordlund (2002), more precisely $n-2=\beta$}.
Combining these expressions with eqn~(\ref{eq_pn2}), they infer
\begin{eqnarray}
m \simeq {\rho_0 L_0^3 \over {\cal M}_{a,0} } \left( { L \over L_0} \right)^{6-n}, 
\label{eq_pn3}
\end{eqnarray}
where $L_0$ is the largest or integral scale of the system and ${\cal M}_{a,0}$
the corresponding Mach number. To get a mass spectrum, it is further assumed 
that the number of cores, $N(L)$, formed by a velocity fluctuation of scale $L$, 
is proportional to $L^{-3}$. Combining this last relation with eqn~(\ref{eq_pn3})
leads to
\begin{eqnarray} 
N(m) d \log m \simeq m^{-3/(6-n)} d \log m.
\label{eq_pn4}
\end{eqnarray}
For a value of $n=3.74$ (close to what is inferred from 3D numerical simulations), 
one gets $N(m) \simeq m^{-1.33}$, very close to the Salpeter exponent.

So far, gravity has not been playing any role in this derivation and the 
mass spectrum that is inferred is valid for arbitrarily small masses.
In a second step, these authors consider a distribution 
of Jeans masses within the clumps induced by turbulence. As the density 
in turbulent flows presents a lognormal distribution, they {\it assume} 
that this implies a lognormal distribution of Jeans lengths and they 
multiply the mass spectrum (\ref{eq_pn4}) by a 
distribution of Jeans masses, which leads to
\begin{eqnarray} 
N(m) d \log m \simeq m^{-3/(6-n)} \left( \int _0^m p(m_J) dm_J \right)  d \log m.
\label{eq_pn5}
\end{eqnarray}
The shape of the mass spectrum stated by eqn~\ref{eq_pn5} is very similar to 
the observed IMF (see for example the figure 1 of Padoan \& Nordlund 2002).

Note, however, that  difficulties with this theory have been pointed out by 
McKee \& Ostriker (2007) and Hennebelle \& Chabrier (2008).
Eqn~(\ref{eq_pn1}), in particular, implies that in the densest regions
where dense cores form, the magnetic field 
is proportional to the density, in strong contrast with what is observed both in simulations
(Padoan \& Nordlund 1999, Hennebelle et al. 2008) and in
observations (e.g. Troland \& Heiles 1986). This is a consequence of 
the assumption that the magnetic field and the velocity field are perpendicular,
which again is not the trend observed in numerical simulations.
In both cases, it is found 
that at densities lower than about 10$^3$ cm$^{-3}$, $B$ depends only weakly 
on $n$ while at higher densities, $B \propto \sqrt{\rho}$. This constitutes a  
problem for this theory, as the index of the power law slope is a direct consequence of 
eqn~(\ref{eq_pn1}). Assuming a different relation between $B$ and $\rho$,  as the aforementioned observed one, would 
lead to a slope stiffer than the Salpeter value. Furthermore, the Salpeter IMF is recovered
in various purely hydrodynamical 
simulations (e.g. Bate et al. 2003), while the Padoan \& Nordlund theory 
predicts a stiffer distribution ($dN/ dM \propto M^{-3}$) in the hydrodynamical case.
Another important shortcoming of this theory is that it predicts that turbulence, by producing overdense, gravitationally unstable areas, {\it always} promotes
star formation, while it is well established 
from numerical simulations that the net effect of turbulence is to 
reduce the star formation efficiency (e.g. MacLow \& Klessen 2004).

\subsection{Turbulent dispersion}
Recently, Hennebelle \& Chabrier (2008, 2009, HC08, HC09) proposed 
a different theory which consists in counting the mass of the fluid regions
within which gravity dominates over the sum of all supports, thermal, turbulent and 
magnetic, according to the Virial condition. In this approach, the role of turbulence
is dual: on one hand it promotes star formation by locally compressing the gas but on the 
other hand, it also quenches star formation because of the turbulent 
dispersion of the flow, which is taken into account in the selection of the pieces of fluid that collapse. 

The  theory is formulated by deriving an extension of the Press \& Schechter (1974, PS) statistical formalism, developed in cosmology.
The two major  differences are (i) the underlying density field, characterized by
small and Gaussian fluctuations in the cosmological case while  lognormal in the 
star formation case, and (ii) the selection criterion, a simple scale-free density 
threshold in cosmology while scale-dependent, based on the Virial theorem in the second case.
That is,  fluid particles which satisfy the criterion (see HC08)
\begin{eqnarray}
  \langle V_{\rm rms}^2\rangle    + 3\, (C_s^{eff})^2 < - E_{\rm pot} / M
\label{viriel_ceff}
\end{eqnarray}
are assumed to collapse and form a prestellar bound core.
The turbulent rms velocity obeys a power-law correlation with the size of the
region, the observed so-called Larson relation,
$\langle V_{\rm rms}^2\rangle =  V_0^2 \times \left( {R \over  1 {\rm pc}} \right) ^{2 \eta}$,
with $V_0\simeq 1\, {\rm km\, s}^{-1}$ and $\eta \simeq 0.4$-0.5 (Larson 1981).

 The principle of the method is the following. First, the density field is smoothed at a scale $R$, using a window function. Then, the total mass contained in  areas which, at scale $R$,  have a density contrast larger than the  specified density  criterion $\delta_R^c$,  is obtained by integrating accordingly the density PDF. 
This mass, on the other hand, is also equal to the total mass located in structures
of mass larger than a scale dependent critical mass $M_R^c$, which will end up forming structures of mass {\it smaller than or equal} to this critical mass for collapse (see HC08 \S5.1). 
\begin{eqnarray}
 \int ^{\infty} _ {\delta_R^c} \bar{\rho} \exp(\delta)   {\cal P}_R(\delta)  d\delta
\label{hc_eq1} = \int _0 ^ {M_R^c} M' \, {\cal N} (M')\,   P(R,M')\, dM'. 
\end{eqnarray}
In this expression, $M_R^c$ is the mass which at scale $R$ is 
gravitationally unstable, $\delta_R^c = log(\rho_R^c/\bar{\rho})$ 
and $\rho_R^c= M_R^c / (C_m R^3)$, $C_m$ being a dimensionless coefficient of 
order unity. ${\cal P}_R$ is the (turbulent) density PDF, assumed to be lognormal, while 
$P(R,M')$ is the conditional probability to find a gravitationally  unstable mass, $M'$
embedded into $M_R^c$ at scale $R$,  assumed to be equal to 1 (see HC08 \S5.1.2 and App. D).
Note that this expression is explicitly solving the cloud in cloud problem 
as the mass which is unstable at scale $R$ is spread over the structures 
of masses {\it smaller} than $M_R^c$ (right hand side integral).  
Therefore, by construction, all the gravitationally unstable regions in the 
parent clump that will eventually  collapse to form individual 
prestellar cores are properly accounted for in this theory (see HC08 \S5.1).

Taking the derivative of 
eqn~(\ref{hc_eq1}) with respect to $R$, we obtain the mass spectrum
\begin{eqnarray}
\label{n_general}
 {\cal N} (M_R^c)  &=& 
 { \bar{\rho} \over M_R^c} 
{dR \over dM_R^c} \,
\left( -{d \delta_R ^c \over dR} \exp(\delta_R^c) {\cal P}_R( \delta_R^c) + \int _  {\delta_R^c}^\infty \exp(\delta) {d {\cal P}_R \over dR} d\delta \right).
\end{eqnarray}
While the second term is important to explain the mass spectrum of unbound clumps defined 
by a uniform density threshold (as the CO clumps), it plays a minor role for 
(Virial defined) bound cores 
 and can generally be dropped (see HC08 for details).

After soma algebra and proper normalisation, one gets
\begin{eqnarray}
{\cal N} (\widetilde{M} ) &=& 2\, {\cal N}_0 \, { 1 \over \widetilde{R}^6} \,
{ 1 + (1 - \eta){\cal M}^2_* \widetilde{R}^{2 \eta} \over
[1 + (2 \eta + 1) {\cal M}^2_* \widetilde{R}^{2 \eta}] }
\times    \left( {\widetilde{M} \over \widetilde{R}^3}  \right) ^{-{3 \over  2} -   {1 \over 2 \sigma^2} \ln (\widetilde{M} / \widetilde{R}^3) }
\times {\exp( -\sigma^2/8 ) \over \sqrt{2 \pi} \sigma },
\label{grav_tot2}
\end{eqnarray}
where $\widetilde{R}= R / \lambda_J^0$, $\delta_R^c = \ln \bigg\{ (1 + {\cal M}^2_* \widetilde{R}^{2 \eta}) / \widetilde{R}^2  \bigg\} $, ${\cal N}_0=  \bar{\rho} / M_J^0$ and $M_J^0$, $\lambda_J^0$ denote the usual thermal Jeans mass and Jeans length, respectively, and
\begin{eqnarray}
\widetilde{M} (R)=  M / M_J^0 = \widetilde{R}\,
(1+ {\cal M}^2_* \widetilde{R}^{2 \eta})
\label{Mtilde}
\end{eqnarray}
denotes the unstable mass at scale $R$ in the turbulent medium.
The theory is controlled by two Mach numbers, namely
\begin{eqnarray}
{\cal M}_* = { 1  \over \sqrt{3} } { V_0  \over C_s}\left({\lambda_J^0 \over   1 {\rm pc} }\right) ^{ \eta}
\approx (0.8-1.0) \,\left({\lambda_J^0\over 0.1\,{\rm pc}}\right)^{\eta}\,\left({C_s\over 0.2 {\rm km \, s^{-1} }}\right)^{-1},
\label{mach_eff}
\end{eqnarray}
defined as the non-thermal velocity to sound speed ratio at the mean   Jeans scale $\lambda_J^0$ (and not at the local Jeans length), and the usual Mach number, ${\cal M}$,
which represents the same quantity at the scale of the turbulence injection scale, $L_i$, assumed to be the characteristic size of the system,
${\mathcal M}={\langle V^2 \rangle^{1/2} \over C_s}$.

 The global Mach number, ${\cal M}$,  broadens the density PDF, as $\sigma^2 = \ln (1 + b^2 {\cal M}^2)$,
illustrating the trend of supersonic turbulence to promote star formation by 
creating new overdense collapsing seeds. 

 The effect described by ${\cal M}_*$  is the  additional non thermal support
induced by the turbulent dispersion.
In particular, at large scales the net effect of turbulence is to stabilize
 pieces of fluid that would be gravitationally unstable if only the thermal 
support was considered. This is illustrated by eqn~(\ref{Mtilde})
which reduces to 
$\widetilde{M} = \widetilde{R}$ when ${\cal M}_*=0$. In particular, for a finite cloud size, 
the gas whose associated turbulent Jeans length is 
larger than the cloud size is not going to collapse.

 When ${\cal M}_* \ll 1$, i.e. the turbulent support is small compared to the thermal one, 
eqn~(\ref{grav_tot2}) shows that
the CMF at large masses  is identical to the Padoan et al. (1997) result, 
i.e. $dN/d\log\, M \propto M^{-2}$. On the other hand, when ${\cal M}_* \simeq 1$,
 $dN/d\log\, M \propto M^{-(n+1)/(2n-4)}$, where the index of the velocity powerspectrum  $n$
is related to $\eta$ by the relation $\eta=(n-3)/2$ (see HC08). As $n \simeq 3.8-3.9$ in 
supersonic turbulence simulations (e.g. Kritsuk et al. 2007),
turbulent dispersion leads to the correct Salpeter slope  and Larson velocity-size relation.

Comparisons with the Chabrier (2003) IMF have been performed for a series of cloud parameters 
(density, size, velocity dispersion) and good agreement has been found 
(HC09) for clouds typically 3 to 5 times denser than the 
mean density inferred from Larson (1981) density-size relation.
Comparisons 
with numerical simulations have also been performed. In particular, 
Schmidt et al. (2010), performing supersonic isothermal simulations with various 
forcing, have computed the mass spectrum of cores supported 
either by pure thermal support or by turbulent plus thermal support.
  Their converged simulations show very good {\it quantitative} agreement 
with the present theory, confirming that turbulent support is needed to 
yield the Salpeter index. Note that Schmidt et al. (2010) use for the density 
PDF the one they measure in their simulations which is nearly, but not exactly lognormal.
Comparisons with the results of SPH simulations 
(Jappsen et al. 2005) including self-gravity and thermal properties of the gas have also
been found to be quite successful (HC09).

\subsection{Difficulties of the gravo-turbulent theories}
One natural question about any IMF theory is to which extent it varies with 
physical conditions. Indeed, there is strong observational support for a nearly invariant form and peak location of the IMF in various environments under Milky Way like conditions (see e.g. Bastian et al. 2010).
Jeans length based theories could
have difficulty with the universality of the peak position, since
 it is linked to the Jeans mass which varies with the gas density. Various propositions
have been made to alleviate this problem. Elmegreen et al. (2008) and Bate (2009) propose
that the gas temperature may indeed increase with density, resulting in a
Jeans mass which weakly depends on the density, while HC08 propose that 
for clumps following Larson relations, there is a compensation between the 
density dependence of the Jeans mass and the Mach number dependence 
of the density PDF, resulting in a peak position that is insensitive to the clump
size.

A related problem is the fact that massive stars are often observed to be located in the 
densest regions, where the Jeans mass is smaller. Indeed, $M_J \propto \rho^{-1/2}$ when a purely thermal support is considered, whereas
$M_J \propto \rho^{-2}$ when turbulence is taken into account (assuming 
that $V \propto L^{0.5}$) (see HC09).
 This constitutes a difficulty for theories based on Jeans mass 
although, as seen above, the issue is much less severe when turbulent support is considered
as massive stars can be formed at densities only few times smaller than the densities
at which low mass stars form. Another possibility is that dynamical interactions 
between young protostars may lead to the migration of massive stars in the 
center of the gravitational well.

Furthermore, the dependence of the freefall time on the Jeans mass
should also modify the link between the CMF and the IMF, as
pointed out by Clark et al. (2007). 
This is particularly true for theories which invoke only  thermal 
support. When  turbulent support is included,
 the free-fall time is found to depend only weakly on the mass, with
$t_{ff}\propto M^{1/4}$ (see McKee \& Tan 2003 and HC09 App. C), 
resolving this collapsing time problem. 
We stress that this time represents 
the time needed for the whole turbulent Jeans mass to be accreted. 
It is certainly true that, within this turbulent Jeans mass, small 
structures induced by turbulent compression will form rapidly. 
Their total mass, however, is expected to represent only a fraction 
of the total turbulent Jeans mass because the net effect of turbulence is 
to decrease the star formation efficiency (see HC08). 

Generally speaking, the fragmentation that occurs during the collapse
could constitute a problem for theories invoking Jeans masses.
Although this problem is far 
from being settled, it should be stressed that such a fragmentation process is not incompatible
with the calculations performed by HC08. As shown by eqn~\ref{hc_eq1}, the presence of small 
self-gravitational condensations induced by turbulence 
 at the early stages of star formation and 
embedded into larger ones is  self-consistently taken into account in the theory. 
Moreover, the SPH simulations performed by Smith et al. (2008) show a 
clear correlation between the initial masses within the gravitational well and 
the final sink masses up to a few local freefall times (see Chabrier \& Hennebelle 2010 for
a quantitative analysis),  suggesting that 
the initial prestellar cores do not fragment into many objects. 
As time goes on, the correlation
becomes weaker but seems to persist up to the end of their run. Massive stars, 
on the other hand, 
are weakly correlated with the mass of the potential well in which they form.
Whether their mass was contained into a larger more massive well 
with which the final sink mass would be well correlated  remains an open issue, which needs to be further 
investigated.
At last, both the magnetic field (e.g. Machida et al. 2005, 
Hennebelle \& Teyssier 2008) and the radiative 
feedback (Bate 2009, Offner et al. 2009) will reduce 
the fragmentation, suggesting that the core-sink correlation found in Smith et al. (2008) should improve if such processes were included.
Clearly, these questions require careful investigations.

\section{Conclusion}
We have reviewed the most recent theories which have been proposed to explain the 
origin of the IMF. Due to limited space, it was not possible to cover 
all of them (e.g. Kunz \& Mouschovias 2009). 
Two main categories received particular attention: the theories
based on accretion and the ones based explicitly on turbulence. It should be 
stressed that these theories are not all exclusive from each other  and may
apply in different ranges of mass. For instance, the turbulent dispersion theory
 calculates the distribution of the initial mass accretion 
reservoirs;  it is not incompatible with the stopped accretion theories and 
with the competitive accretion as long as mass redistribution/competition  occurs within one 
parent core reservoir. 
The question as to whether one of these mechanisms is dominant is yet unsettled. 
Detailed comparisons between systematic sets of simulations, as done in Schmidt et al. (2010), or observations, and the various analytical 
predictions is clearly mandatory to make further progress.

\end{document}